\documentclass{elsart}  %% ,twocolumn}
\usepackage{amssymb,amsmath,multirow,epsfig,graphicx,color}

\usepackage{epsfig}
\usepackage{graphicx,amsmath}
\usepackage{color}
\newcommand\ba{\begin{eqnarray}}
\newcommand\ea{\end{eqnarray}}

\newcommand{\be}{\begin{equation}}
\newcommand{\ee}{\end{equation}}
\newcommand{\bas}{\begin{eqnarray*}}
\newcommand{\eas}{\end{eqnarray*}}

\def\sla#1{\rlap\slash #1}

\newcommand{\cc}{\c c}

\newcommand{\bno}{\begin{eqnarray*}}
\newcommand{\eno}{\end{eqnarray*}}
\newcommand{\nl}{\newline}

\def\qbar{\bar{q}}

\def\auj{\number\day\space\ifcase\month\or
janvier\or f\' evrier\or mars\or avril
\or mai\or juin\orjuillet\or ao\^ut
\or septembre\or octobre\or novembre
\or d\' ecembre\fi\space\number\year}

\def\hoje{\number\day\space de \ifcase\month\or
Janeiro,\or Fevereiro,\or Mar\cc o,\or 
Abril,\or Maio,\or Junho,\or Julho,
\or Agosto,\or Setembro,\or Outubro,\or Novembro,
\or Dezembro,\fi\space\number\year}

\def\date{\number\day\space  \ifcase\month\or
January,\or February,\or March,\or 
Appril,\or May,\or June,\or July,
\or Agust,\or septembre,\or Octubre,\or November,
\or December,\fi\space\number\year}

\begin{document}

\begin{frontmatter}

%% \title{
\vspace{-1cm} 
\begin{flushright}
    {\large \bf LFTC-18-4/25 \vspace{1cm}}
\end{flushright}
\title{ 
$\rho$-meson properties in medium
}

\author[UNICSUL]{J. P. B. C. de Melo},
%\email[E-mail: ]{joao.mello@cruzeirodosul.edu.br}
\author[UNICSUL]{K. Tsushima}
%% , and
%% \author[ITA]{ T. Frederico}
%\email[E-mail: ]{tobias@ita.br}
\address[UNICSUL]{Laborat\'orio de F\'\i sica Te\'orica e 
Computacional - LFTC, Universidade Cruzeiro do Sul, 01506-000 S\~ao Paulo, Brazil} 
%% \address[ITA]{Instituto Tecnol\'ogico de Aeron\'autica, 
%% DCTA, 12.228-900 S\~ao Jos\'e dos Campos, Brazil}
%%\date{\today}
%%  \date
%\pacs{13.66.Bc, 12.20.-m,13.40.-f,13.88.+e}

%\vspace{0.5cm}
\begin{abstract}
Properties of $\rho$-meson in symmetric nuclear matter are investigated in a
light-front constituent quark model (LFCQM), using the in-medium inputs 
calculated by the quark-meson coupling (QMC) model. 
The LFCQM used in this study was already applied for the studies of   
the electromagnetic properties of $\rho$-meson in vacuum, namely, 
the charge~$G_0$, magnetic~$G_1$, and quadrupole~$G_2$ form factors,   
electromagnetic charge radius, and electromagnetic decay constant.
Using the two different density dependence of the regulator mass in medium, 
we predict that the charge radius, and quadrupole moment are enhanced 
as increasing the nuclear matter density, while the magnetic moment is slightly quenched.
Furthermore, we predict the value $Q^2_{\rm zero}$,  
which crosses zero of the charge form factor, $G_0(Q^2_{\rm zero})=0$ 
($Q^2 = -q^2  > 0$ with $q$ being the four-momentum transfer),  
decreases as increasing the nuclear matter density 
by the two different density dependence of the regulator mass.
On the other hand, for the electromagnetic decay constant of the $\rho$-meson, 
the two different density dependence of the regulator mass predict 
the opposite density dependence. 
Namely, as increasing the nuclear matter density, 
the naive treatment with the density independent regulator mass as in the vacuum, 
predicts the increase of the decay constant, 
while the other that assumes the same density dependence of the regulator mass  
as that of the in-medium constituent quark mass, 
predicts the decrease of the decay constant.
Thus, although the other physical quantities are predicted to have similar 
density dependence by the two different density dependence of the regulator 
mass applied, the density dependence of the $\rho$-meson electromagnetic 
decay constant are predicted to have opposite density dependence, 
and the facts suggest that the in-medium $\rho$-meson decay constant 
needs to be investigated further in the future. 
\end{abstract}


\end{frontmatter}

\section{Introduction}

One of the fundamental objectives in hadronic physics is to understand 
the structure of hadrons in terms of the quark and gluon degrees of freedom,   
the basis of quantum chromodynamics~(QCD).   
The Standard Model~(SM) of elementary particles contains QCD as 
the strong interaction theory, and practicing QCD 
to understand the hadron structure is an important part of understanding SM. 
However, it is not straightforward to apply QCD directly to study the 
properties of hadrons such as mesons, baryons, and tetra-quarks, 
the bound state systems of quarks and gluons, in particular in the low energy 
nonperturbative region~(see Refs.~\cite{Ydarin,Skands2012}).  
Despite many successes of QCD which is believed as the correct 
quantum field theory of strong interaction~\cite{Brosky1998,Vary2010}, 
the hadronic properties in the low-energy region  
cannot be directly extracted naively. 
To overcome the difficulties, effective treatments of QCD, such as 
constituent quark models~(CQM) and light-front treatment of hadrons  
have been developed, and achieved impressive success~[5-27]. 
%% \cite{Dziembowski87,Pacheco99,Simula2002,Melo2002,Melo2003,Melo2004,Hwang2003,Huang2004,Braguta2004,Salcedo2004,
%%Salcedo2006,Karmanov2007,Pacheco2002,Melo2006,Tsirova2009,Ji2001,
%%Choi2001,Pena2014,Yabusaki2015,deAraujo:2017uad,Horn2016,Vary2016,Bacchetta2017}.
In particular, we emphasize the cases of spin-1 vector 
particles~[28-46], 
%%\cite{Cardarelli1995,Melo2012,Pacheco992,Lev1999,Lev2000,Jaus2003,Aliev04,Melo1997,
%%Choi2004,Bhagwat2008,Gudino2015,Simonis2016,Serrano2015,Krutov2016,Roberts2011,Biernat2014,Choi2014,Dong2017,deMelo2018b2}, 
that are of relevant for the present 
study of $\rho$-meson.

For the $\rho$-meson, experimental data are very scarce at present. 
In Refs.~\cite{Huber1998,Huber2003} some properties of the $\rho$-meson 
in medium were discussed based on the experiment data.  
Recently, in Refs.~\cite{Adamuscin2007,deMelo2016} the data from 
BaBar Collaboration~\cite{Aubert2009}   
for the $e^+ e^- \to \rho^+ \rho^-$ reaction in vacuum 
were analyzed to study the electromagnetic properties of 
$\rho$-meson based on perturbative QCD.
%%
%% Choi2015b,Roberts,Thomas,Fanelli2016}, also, that approach, 
%% is very closely related with the experimental data~\cite{PDG2014}. 

Light-front quantum field theory~(LFQFT) on which the present study bases, 
is able to incorporate the following two important aspects simultaneously,  
namely, QCD and the picture of constituent quark model. 
It is a natural approach to calculate physical observables 
based on the quark degrees of freedom~\cite{Brosky1998,Vary2010}. 
In the light-front approach one can have relativistic wave functions   
of hadronic bound states described in terms of quarks and gluons, 
and the approach is suitable for understanding the hadron substructure 
focusing on nonperturbative aspect~\cite{Cardarelli1995,deMelo2016,Jaus1990,
Jaus1991,Choi2015b,Choi2015,Fanelli2016}. 
%% 
%% With Light-front quantum field theory, it is possible, describe 
%% the lower Fock componets of the hadronic wave function 
%% bound states, i.e, mesons and baryons in terms of 
%% quarks and gluons~\cite{Brosky1998,Vary2010, Melo2006}.

Based on the advantages mentioned above for the light-front approach, 
we use a light-front constituent quark model (LFCQM) 
which was already applied for the studies of $\rho$-meson electromagnetic 
properties in vacuum~\cite{Melo1997,Melo2012b,Anace2014}.    
We extend the preliminary study made in symmetric nuclear matter  
for the $\rho$-meson properties in medium~\cite{deMelo:2016ynt}, 
and elaborate in the present work. 
In this work we use a different regularization treatment for the calculation 
in nuclear medium. Namely, we assume the same density dependence of the regularization 
mass as that of the in-medium constituent quark mass. 
By the use of this density dependent regularization mass, 
we calculate the $\rho$-meson electromagnetic  
charge~$G_0$, magnetic~$G_1$, and quadrupole~$G_2$ form factors, 
electromagnetic square charge radius~$r^2_{\rho}$, and   
$\rho^0 \to e^+ e^-$ decay constant $f_{\rho}$ in symmetric nuclear matter.
The results obtained using the density independent regularization mass 
were presented in Ref.~\cite{deMelo:2016ynt}.
Then, the results of the two different regularization treatments 
applied in medium are compared and discussed with those of Ref.~\cite{deMelo:2016ynt}. 
However, we emphasize that, except for the density dependence of 
the $\rho$-meson decay constant, the results of two different regularization treatments 
predict the similar density dependence of the $\rho$-meson in-medium electromagnetic properties. 
Since the density dependence of the $\rho$-meson decay constant 
shows the opposite density dependence, 
this needs to be studied further by using other model, and/or developing  
a proper in-medium regularization method in the future.

We remind that, we use the plus-component of the 
electromagnetic current $J_\rho^+$ in this study, 
considering the situation that the $\rho$-meson 
is immersed in symmetric nuclear matter~\cite{Hayano:2008vn,Brooks2011,Melo2014}. 
(For comprehensive reviews on hadronic and quark properties in nuclear medium,  
e.g., see Refs.~\cite{Hayano:2008vn,Brooks2011}.) 
The model for the $\rho$-meson in vacuum we use~\cite{Melo1997} in this study, 
is constrained by the ``angular condition'' for the light-front electromagnetic 
current matrix elements~\cite{Cardarelli1995,Melo2012,Melo1997,Melo2012b,Anace2014,Melo2004bjp}.
Any light-front-based models of spin-1 particles should satisfy 
the angular condition, which was originally discussed in Ref.~\cite{Inna84}.

\section{Quark-meson coupling model}
\label{QMC}

To describe symmetric nuclear matter (many nucleon system), 
we rely on the quark-meson coupling~(QMC) model, which bases on 
the quark degrees of freedom, and calculate necessary in-medium 
inputs for the quarks and hadrons to implement in the light-front 
constituent quark model, similarly to the studies made  
for pion and kaon~\cite{Melo2014,Melo2016,Melo2017,Yabusaki:2017sgs}  
and nucleon~\cite{deAraujo:2017uad}. 
Detail of the light-front constituent quark model we use in this study   
is described in Refs.~\cite{Melo2002,Melo2003,Yabusaki2015}  
(and for nucleon in Ref.~\cite{deAraujo:2017uad}).

The QMC model was invented by Guichon~\cite{Guichon1988} using the MIT bag model, 
and Frederico et al.~\cite{Frederico1989} using a relativistic confining harmonic potential.
The model was successfully applied for studying the properties of 
finite nuclei~\cite{QMCfinite}, and the properties of hadrons 
in medium~\cite{Saito2007,Krein:2017usp}. 
(See Ref.~\cite{Saito2007} for other approaches similar to the QMC model used in the present study.)
In the QMC model the meson and baryon internal structure is modified in medium 
due to the surrounding medium, by the self-consistent exchange of 
the scalar-isoscalar ($\sigma$), vector-isoscalar ($\omega$), 
and vector-isovector ($\rho$) meson fields 
directly coupled to the relativistic, confined light-quarks in the nucleon (hadron), 
rather than to the point-like nucleon (hadron).
We briefly explain next the main feature of the QMC model, which 
is used to calculate the in-medium inputs necessary to study 
the $\rho$-meson electromagnetic properties in symmetric nuclear matter.
 
We consider a system of infinite, uniform, spin and isospin saturated symmetric nuclear matter 
in the Hartree mean field approximation in the rest frame of matter. 
Thus, irrelevant $\rho$-meson filed is suppressed in the following, 
since the total isospin of the matter is zero, and in the Hartree approximation $\rho$-meson 
mean filed becomes zero.
In addition quantities with an asterisk, $^*$,  will stand for those in medium hereafter. 
The effective Lagrangian density of the QMC model at the hadron level 
may be given by~\cite{Saito2007},
\begin{equation}
{\cal L} = {\bar \psi} [i\gamma \cdot 
\partial -m_N^*({\hat \sigma}) -g_\omega {\hat \omega}^\mu \gamma_\mu ] \psi
+ {\cal L}_\textrm{meson},
\label{lag1}
\end{equation}
with ${\cal L}_\textrm{meson}$ being the free meson Lagrangian,   
\begin{equation}
{\cal L}_\mathrm{meson} = \frac{1}{2} 
(\partial_\mu {\hat \sigma} \partial^\mu {\hat \sigma} - m_\sigma^2 {\hat \sigma}^2)
- \frac{1}{2} \partial_\mu {\hat \omega}_\nu (\partial^\mu {\hat \omega}^\nu - \partial^\nu {\hat \omega}^\mu)
+ \frac{1}{2} m_\omega^2 {\hat \omega}^\mu {\hat \omega}_\mu \ . \nonumber
\label{mlag1}
\end{equation}
In the above $\psi$, ${\hat \sigma}$ and ${\hat \omega}$ are respectively the nucleon,
Lorentz-scalar-isoscalar $\sigma$, and Lorentz-vector-isoscalar $\omega$ 
field operators, with $g_{\omega}$ being the nucleon-$\omega$~coupling constant, 
while the nucleon-$\sigma$ effective coupling which depends on the $\hat{\sigma}$ 
(or nuclear density) is defined by, 
\begin{equation}
m_N^*({\hat \sigma}) = m_N - g_\sigma({\hat \sigma}) {\hat \sigma}, 
\label{efnmas}
\end{equation}
where the effective nucleon mass is denoted by $m_N^*$.
Then, the nucleon density $\rho_N$,  
the nucleon Fermi momentum $k_F$, the nucleon scalar density $\rho_s$, and the  
effective nucleon mass $m^*_N$ are related by,
\begin{eqnarray}
\rho_N &=& \frac{4}{(2\pi)^3}\int d^{\,3}k\ \theta (k_F - |{\bf k}|)
= \frac{2 k_F^3}{3\pi^2},
\label{rhoB} \nonumber \\ 
\rho_s &=& \frac{4}{(2\pi)^3}\int d^{\,3}k \ \theta (k_F - |{\bf k}|)
\frac{m_N^*(\sigma)}{\sqrt{m_N^{* 2}(\sigma)+{\bf k}^2}}, 
\label{rhos}
\end{eqnarray}
where~$m^*_N(\sigma)$ is the value of the effective nucleon mass at a given  
density, calculated by the QMC model~\cite{QMCfinite,Saito2007}. 

The Dirac equations for the light quarks and light antiquarks in the bag of 
hadron $h$ in nuclear matter at the position $x = (t,{\bf r})$ 
with $|{\bf r}| \le$ bag radius, are given by~\cite{Saito2007},
\begin{eqnarray}
\left[ i \gamma \cdot \partial_x -
(m_q - V^q_\sigma)
\mp \gamma^0
\left( V^q_\omega +
\frac{1}{2} V^q_\rho
\right) \right]
\left( \begin{array}{c} \psi_u(x)  \\
\psi_{\bar{u}}(x) \\ \end{array} \right) &=& 0,
\label{diracu} \nonumber \\
\left[ i \gamma \cdot \partial_x -
(m_q - V^q_\sigma)
\mp \gamma^0
\left( V^q_\omega -
\frac{1}{2} V^q_\rho
\right) \right]
\left( \begin{array}{c} \psi_d(x)  \\
\psi_{\bar{d}}(x) \\ \end{array} \right) &=& 0,
\label{diracd}
%% 
% \left[ i \gamma \cdot \partial_x - m_{Q} \right]
%%\psi_{Q} (x)\,\, ({\rm or}\,\, \psi_{\Qbar}(x)) &=& 0,  \label{diracQ}
\end{eqnarray}
where, $V^q_\sigma = g^q_\sigma \sigma, V^q_\omega = g^q_\omega \omega$, 
and $V^q_\rho = g^q_\rho b$ are respectively the constant mean filed potentials with 
the corresponding quark-meson coupling constants, $g_\sigma, g_\omega$ and $g^q_\rho$, 
and the Coulomb interactions are neglected, because the nuclear matter 
is described in the strong interaction. 
The vector meson mean fields appearing in $V^q_\omega$ and $V^q_\rho$,  
correspond respectively to the expectation values evaluated in symmetric nuclear matter are, 
$\omega^\mu = (\omega, {\bf 0})$ and 
$\rho^\mu_i = (\delta_{i,3}\, b, {\bf 0})$.
In addition SU(2) symmetry for the light-quark masses, $m_q=m_{\bar{q}} = m_u = m_d$, is assumed.

The normalized, static solution for the ground state light quark ($q$) 
and light antiquark ($\bar{q}$) in the hadron $h$ may be written as 
$\psi_{q,\bar{q}} (x) 
= N_{q,\bar{q}} \, e^{{-i\epsilon_{q,\bar{q}}\, t/{R^*_h}}}\, \psi_{q,\bar{q}} \psi({\bf r})$, 
where $N_{q,\bar{q}}$ and $\psi_{q,\bar{q}}({\bf r})$ are the normalization factor and 
corresponding spin and spatial part of the wave function.
The bag radius in medium of the hadron $h$, $R_h^*$,
is determined by the stability condition for the mass of the hadron 
against the variation of the bag radius~\cite{Saito2007}, to be shown in Eq,~(\ref{hmass}).
The eigenenergies in units of $1/{R^*_h}$ are given by,
%%%%%%%%%%%%%%%%
\begin{eqnarray}
\left( \begin{array}{c}
\epsilon_u \\
\epsilon_{\bar{u}}
\end{array} \right)
& = &  \Omega_q^* \pm R_h^* \left(
V^q_\omega
+ \frac{1}{2} V^q_\rho \right),    \nonumber \\
\left( \begin{array}{c} \epsilon_d \\
\epsilon_{\bar{d}}
\end{array} \right)
& = &  \Omega_q^* \pm R_h^* \left(
V^q_\omega
- \frac{1}{2} V^q_\rho \right), 
\label{energy}
%
%\epsilon_{Q}
%= \epsilon_{\Qbar} =
%\Omega_{Q}.
%\label{energy}
\end{eqnarray}
%%%%%%%%%%%%%%%
where,~$\Omega_q^*=\Omega_{\bar{q}}^*
=[x_q^2 + (R_h^* m_q^*)^2]^{1/2}$, with
$m_q^*=m_q{-}g^q_\sigma \sigma$,
and $x_{q}$ being the lowest mode bag eigenvalue. 
Because we consider symmetric nuclear matter with the Hartree approximation,
$V^q_\rho$ is zero also at the quark level, thus we will ignore hereafter.

The mass of the low-lying hadron $h$ in symmetric nuclear matter, is calculated   
with the bag radius stability condition, 
%%%%%%%%%%%%%%%%
\begin{eqnarray}
m_h^* &=& \sum_{j=q,\bar{q}}
\frac{ n_j\Omega_j^* - z_h}{R_h^*}
+ \frac{4}{3} \pi R_h^{* 3} B,\quad
\left. \frac{\partial m_h^*}
{\partial R_h}\right|_{R_h = R_h^*} = 0,
\label{hmass}
\end{eqnarray}
%%%%%%%%%%%%%%
with $n_q (n_{\qbar})$ being the light-quark (light-antiquark) number. %%  $q$ and $Q$, respectively.

Now we study the $\rho$-meson electromagnetic properties in symmetric nuclear matter 
using the inputs calculated by the QMC model.
To do so, we must rely on the $\rho$-meson model in vacuum which is 
successful and simple enough to handle in extracting the main in-medium properties.
As already mentioned, we use the light-front constituent quark model 
for the $\rho$-meson developed in Ref.~\cite{Melo1997}.

The model uses the light-quark constituent quark mass value in vacuum 
$m_q = m_{\bar{q}} = 430$ MeV. Using this value we calculate  
the corresponding symmetric nuclear matter properties with the QMC model. 
By fitting the nuclear matter saturation properties, namely the binding energy of 15.7 MeV 
at the saturation density $\rho_0 = 0.15$ fm$^{-3}$, we obtain the corresponding 
quark-meson coupling constants. The coupling constants, and some quantities 
calculated in the QMC model at $\rho_0$ are listed in Table~\ref{Tab:QMC}.
For a comparison, the same quantities obtained in the standard QMC model 
with $m_q = 5$ MeV, are also listed in Table~\ref{Tab:QMC} as already mentioned.

%%%%%%%%%%%%%%%%%%
\begin{table}[htb]
\vspace{5ex}
\begin{center}
\caption{The QMC model (MIT bag model) quantities~(see Ref.~\cite{Saito2007} for details),  
coupling constants, the parameters $z_N$ ($z_\rho$)    
for nucleon ($\rho$-meson) [see Eq.~(\ref{hmass})], 
bag constant $B$ ($B^{1/4}$ in [MeV]), 
and some properties for symmetric nuclear matter
at normal nuclear matter density $\rho_0 = 0.15$ fm$^{-3}$,
for $m_q = 5$ and $430$ MeV (the latter is relevant for this study). 
The effective nucleon (quark) mass, $m_N^*$ ($m_q^*$), 
and the nuclear incompressibility, $K$, are quoted in [MeV]. 
The free nucleon bag radius is the input with $R_N = 0.8$ fm,
the standard value in the QMC model~\cite{Saito2007}. 
The vacuum mass value (input) for the $\rho$-meson is $m_\rho = 770$ MeV.
}
\label{Tab:QMC}
\bigskip
\begin{tabular}{c|cccccccc}
\hline
$m_q$ (MeV) &$g_{\sigma}^2/4\pi$&$g_{\omega}^2/4\pi$
&$m_N^*$ & $m_q^*$ &$K$ & $z_N$ & $z_\rho$ & $B^{1/4}$ \\
\hline
 5   &~5.39  &~5.30   &~754.6  &~-135.6 &~279.3  &~3.295 &~1.907 &~170.0 \\
 %220 &~6.40  &~7.57   &~698.6  &~320.9  &~4.327 &~148  &~???  \\
 430 &~8.73  &~11.94  &~565.3 &~245.7 &~361.4  &~5.497 &~2.939 &~69.8  \\
\hline
\hline
\end{tabular}
\end{center}
\vspace{3ex}
\end{table}
%\vspace*{3cm}
%%%%%%%%%%%%%%

\begin{center}
\begin{figure}[ht]
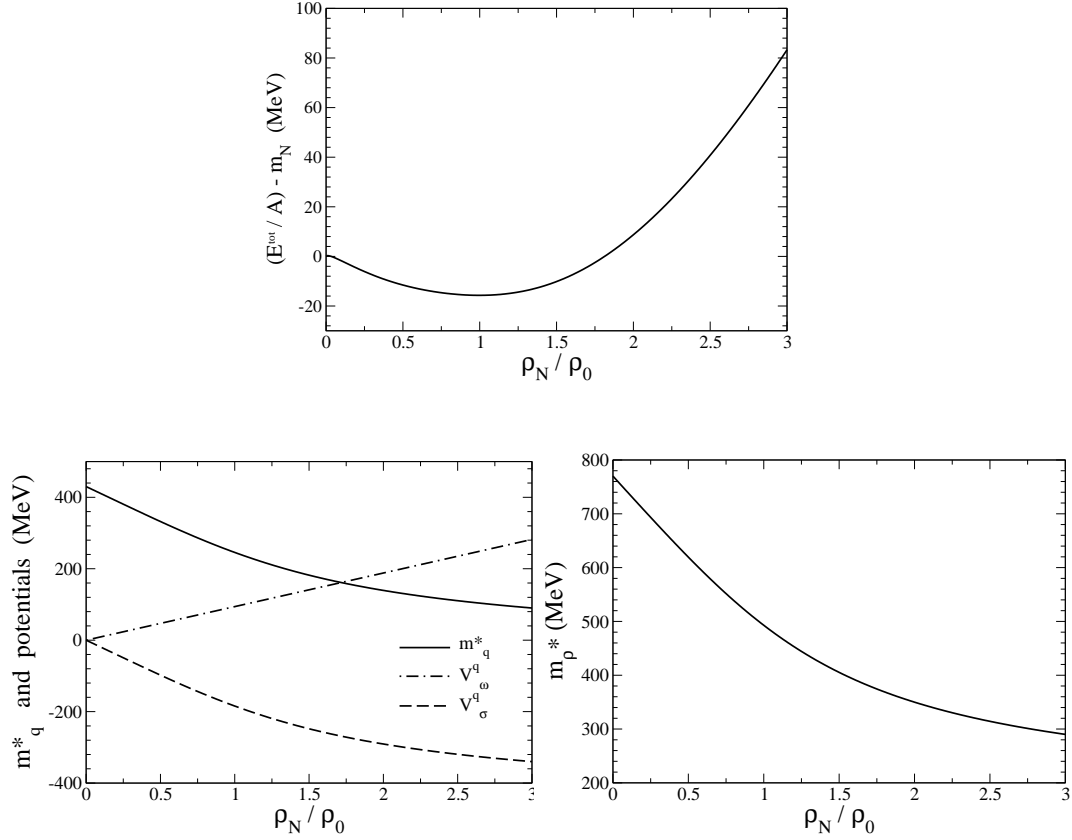

\begin{center}
\vspace{3ex}
\epsfig{figure=LFrhoEnergyDen.eps,width=7.0cm}
\vspace{5ex}
\end{center}
\epsfig{figure=LFrho_mqstar.eps,width=7.0cm}
\epsfig{figure=LFrho_rhomstar.eps,width=7.0cm}
\caption{\label{fig:qmc}
Energy per nucleon $(E^{\rm Total}/A) - m_N$ (upper panel),  
effective quark masses and the potentials felt by the light quarks (lower-left panel), 
and effective $\rho$-meson mass (lower-right panel) 
in symmetric nuclear matter calculated 
in the QMC model~\cite{Saito2007}.
}
\end{figure}
\end{center}

One of the noticeable differences among the quantities calculated with the 
standard value $m_q = 5$ MeV and those with the value $m_q = 430$ MeV, 
is the nuclear incompressibility, $K$. 
It yields a larger value of $K = 361.4$ MeV with $m_q = 430$ MeV, 
while $K = 279.3$ MeV with $m_q = 5$ MeV.
The corresponding energy density per nucleon for $m_q = 430$ MeV, 
$(E^{\rm Total}/A) - m_N$, is shown in Fig.~\ref{fig:qmc} upper panel.

%%%%%%%%%%%%%%%%%%%
%%%%%%%%%%%%

Concerning the solid line shown in Fig.~\ref{fig:qmc} upper panel, 
the curvature of the energy density 
versus $\rho/\rho_0$ is larger than that for $m_q = 5$ MeV. 
Thus $(E^{\rm Total}/A) - m_N$ varies faster than that for $m_q = 5$ MeV 
as the nuclear matter density increases. 
(See Ref.~\cite{Saito2007} for the curve of $(E^{\rm Total}/A) - m_N$ with $m_q = 5$ MeV.) 
Next, we show also in Fig.~\ref{fig:qmc} lower-left panel 
the effective light-quark mass $m^*_q = m_q - g^q_\sigma \sigma$,   
scalar potential $V^q_\sigma = g^q_\sigma \sigma$, and vector potential 
$V^q_\omega = g^q_\omega \omega$ felt by the light quarks, 
and in lower-right panel the effective $\rho$-meson mass $m^*_\rho$. 
(See Eq.~(\ref{hmass}) for $m^*_\rho$ with $h \to \rho$.)
Note that, the $\rho$-meson effective mass $m^*_\rho$ shown in 
Fig.~\ref{fig:qmc} lower-right panel,   
corresponds to that of naive SU(6) quark models, and readers must not be confused 
with that of the $\rho$-meson mean field (operator) appearing in the QMC model.
In addition we have neglected the width of the $\rho$-meson, following the 
usual practices of naive SU(6) quark models.

%\newpage
\section{$\rho$-meson electromagnetic form factors} 
\label{rhoEMFFs}

A general expression of the electromagnetic current 
for a $\rho$-meson (spin-1 particle) is given by~\cite{Gilman:2001yh}:
\begin{equation}
 J^{\mu}_{\alpha \beta} = 
 - \left[ F_1(Q^2) g_{\alpha \beta} - F_3(Q^2) \frac{q_\alpha q_\beta}{2 m^2_{\rho}}  \right] P^{\mu} - 
 F_M(Q^2) \left(q_{\alpha} g^{\mu}_{\beta} - q_{\beta} g^{\mu}_{\alpha} \right), 
 \label{current1}
\end{equation}
where~$m_\rho$ is the mass of the $\rho$-meson, $q^\mu$ the four-momentum transfer 
with $Q^2 = -q^2 > 0$, and $P^{\mu} \equiv (p_i + p_f)^\mu$, 
the sum of the initial ($p_i$) and final ($p_f$) momenta. 
%% The frame utilized here, is Breit frame,~$(0,q_x,0,0)$, and the spin-1 particle polarization. 
%% , with  the Cartezian basis is 
The electromagnetic current, 
$J_{ji}^\mu \equiv \epsilon^{'\alpha}_j J_{\alpha \beta}^+ \epsilon_i^{\beta}$,   
in an impulse approximation is given by,  
\begin{eqnarray}
J^\mu_{ji}&=&\imath  \int\frac{d^4k}{(2\pi)^4}
 \frac{Tr[ \epsilon^{'\alpha}_j \Gamma_{\alpha}(k,k-p_f)
(\sla{k}-\sla{p_f} +m_q) \gamma^\mu 
(\sla{k}-\sla{p_i}+m_q)}
{((k-p_i)^2 - m_q^2+\imath\epsilon) 
(k^2 - m_q^2+\imath \epsilon)
((k-p_f)^2 - m_q^2+\imath \epsilon) } \nonumber \\ 
& &  \epsilon^\beta_i 
(\sla{k}+m_q) \Gamma_{\beta}(k,k-p_i)]  
\times \Lambda(k,p_f)\Lambda(k,p_i) \ ,
\label{electro}
\end{eqnarray}
where  $m_q$ is the quark mass, and without loss of generality, 
the four-momenta of the initial and final states of the $\rho$-meson 
may respectively be chosen by~$p_i^\mu=(p^0,-q/2,0,0)$ and~$p_f^\mu=(p^0,q/2,0,0)$  
with the four-momentum transfer defined by $q^\mu=(0,q,0,0)$ 
to satisfy the Drell-Yan condition~\cite{Melo2012,Melo1997}.
In Eq.~(\ref{electro}) ~$\epsilon^{'\alpha}_j$ and~$\epsilon^{\beta}_i$ are the 
final- and initial-state $\rho$-meson polarization vectors, respectively given by,     
\begin{equation}
\epsilon^{\prime \mu }_x=(\sqrt{\eta},\sqrt{1+\eta},0,0),
~\epsilon^{\prime \mu}_y=(0,0,1,0),
~\epsilon^{\prime \mu }_z=(0,0,0,1),
\end{equation}
and 
\begin{equation}
\epsilon^{\mu}_x=(-\sqrt{\eta},\sqrt{1+\eta},0,0),
~\epsilon^{\mu}_y=(0,0,1,0),~\epsilon^{\mu}_z=(0,0,0,1), 
\end{equation}
with~$\eta = Q^2/m^2_{\rho}$.
The electromagnetic current,~Eq.~(\ref{electro}), is divergent, and in order to 
make $J^\mu_{ji}$ finite, a regulator function~$\Lambda(k,p)$ is used~\cite{Melo1997}: 
\begin{equation}
\Lambda(k,p)=\frac{1}{((k-p)^2 - M^2_R + \imath \epsilon)^2}.
\end{equation}
Here, the regulator mass value $m_R$ to yield $M_R^2$ (see Eq.~(\ref{MR}) for 
$m_R$, and the $M_R^2$ mass operator definition),  
is chosen to reproduce the experimental value 
of the $\rho$-meson decay constant $f_\rho$ (see Eq.~(\ref{fdecay}) for the definition)  
extracted from the $\rho^0 \to e^+ e^-$ decay 
width~\cite{PDG2014}. 
The $\rho$-$q\bar{q}$ vertex with spinor structure is modeled by~\cite{Melo1997},
\begin{equation}
\Gamma^\mu (k,k') = \gamma^\mu -\frac{m_\rho}{2}
 \frac{k^\mu+k'^\mu}{ p \cdot k + m_\rho m_q -\imath \epsilon}, 
\label{eq:rhov}
\end{equation}  
where the $\rho$-meson is on-mass-shell, and its four momentum
is $p^\mu \ = \ k^\mu \ - \  k'^\mu$ with the quark momenta
$k^\mu$ and $k'^\mu$~\cite{Melo2012,Melo1997}.

Working with the light-front coordinate,
$a^\mu=(a^+=a^0+a^3,a^-=a^0-a^3, \vec{a}_{\perp}= (a^1, a^2))$, 
the light-front $\rho$-meson wave function is obtained,  
after substituting with the on-mass-shell condition    
$k^-=(k^2_{\perp}+m_q^2)/k^+$ in the quark propagator 
(see Ref.~\cite{Melo1997} for details), 
\begin{eqnarray}
\Phi_i(x,\vec k_\perp)=\frac{N^2}{(1-x)^2(m^2_\rho-M_0^2)
(m^2_\rho- M^2_R)^2} 
\vec \epsilon_i . [\vec \gamma -  \frac{\vec k}{\frac{M_0}{2}+ m_q}] \ ,
\label{eq:npwf}
\end{eqnarray}
where, $x=k^+/p^+$. The polarization state is given by $\vec{\epsilon}_i$. 
The wave function corresponds to an S-wave state~\cite{Jaus1990}.  
The square of the free mass operator $M_0^2$, 
and the regulator mass operator $M_R^2$, are given by: 
\begin{eqnarray}
M^2_0 & = &  \frac{k^2_\perp+m_q^2}{x} 
+\frac{(\vec p-\vec k)^2_\perp+m_q^2}{1-x}- \vec{p}_\perp^{\, 2}, 
\label{M0}\\ 
M^2_R & = &  \frac{k^2_\perp+m_q^2}{x} 
+\frac{(\vec{p} -\vec{k})_\perp^{\, 2} + m_R^2}{1-x}- \vec{p}_\perp^{\, 2}~.
\label{MR}
\end{eqnarray}

\subsection{Angular condition and electromagnetic form factors}

For the spin-1 particles in the light-front approach, 
matrix elements of the plus-component of electromagnetic current,
$J^+$, is constrained by the angular condition equation 
with the light-front spin basis~~\cite{Cardarelli1995,Melo1997,Inna84}:  
\begin{eqnarray}
 \Delta(q^2 = -Q^2)= ( 1 + 2 \eta ) I^+_{11} + I^+_{1-1} + 
 - \sqrt{8 \eta} I^+_{10}-I^+_{00}=0~.
\label{ang}
\end{eqnarray}
The relations among the light-front basis $I^+_{m'm}$ $(m',m = \pm 1, 0)$ and those of the 
instant form spin basis $J^+_{ji}$ $(j,i = x,y,z)$, can be made by the 
Melosh rotation matrix. 
(See Refs.~\cite{Melo2012,Melo1997} for details.)

With the angular condition Eq.~(\ref{ang}), 
it is possible to arrange the electromagnetic form factors 
to form with different linear combinations~\cite{Cardarelli1995,Melo1997}  
by eliminating some matrix elements~$I^+_{m'm}$. 
However, some linear combinations 
break the covariance as well as the rotational symmetry. 
This is due to the zero mode contributions, or pair term  
contributions~\cite{Melo2012,Bakker2002,Clayton2015}.  
In Ref.~\cite{Melo2012} a careful analysis was made  
for the origins of the zero-mode contributions for the 
matrix elements of the electromagnetic current of spin-1 particles, 
in particular for the $\rho$-meson. 

It was demonstrated that the zero mode contributions are canceled out 
in the combinations of the electromagnetic current matrix 
elements of Grach et al.~\cite{Inna84}, by numerically in Ref.~\cite{Melo1997}, 
and by analytically in Ref.~\cite{Melo2012}. 
The reason is that the electromagnetic matrix element of the current, $I^+_{00}$, 
was eliminated by the angular condition~\cite{Cardarelli1995,Melo1997,Choi2004,Inna84}. 
For some prescriptions in the literature, the zero-mode or 
non-valence contributions needed to be added  
in order to recover the full covariance~\cite{Melo2006,Melo2012,Melo1997,Bakker2002}.

With the prescription of Ref.~\cite{Inna84}, 
the electromagnetic form factors of $\rho$-meson are   
given by both in the light-front spin basis $I^+_{mm'}$ 
and the instant form spin basis $J^+_{ji}$: 
\begin{eqnarray}
G_0 & = & \frac{1}{3}\left[\left( 3 - 2 \eta \right)
 I^+_{11} + 2 \sqrt{2 \eta} I^+_{11} + I^+_{1-1} \right] 
 =  \frac{1}{3}\left[ J^+_{xx} + 2 J^+_{yy} - \eta J^+_{yy}  
 + \eta J^+_{zz}\right], 
\nonumber \\
 G_1 & = & 2 \left[ I^+_{11} -\frac{1}{\sqrt{2 \eta } } I^+_{10}  \right] 
 = \left[  J^+_{yy} - J^+_{zz} -\frac{J^+_{zx}}{\sqrt{\eta}}  \right], 
\nonumber \\
 G_2 & = & \frac{2 \sqrt{2}}{3} \left[ -\eta I^+_{11} + \sqrt{2 \eta} 
 I^0_{10} -I^+_{1-1}  \right] 
 =  \frac{\sqrt{2}}{3} \left[ J^+_{xx} - 
 (1 + \eta) J^+_{yy} + \eta J^+_{zz} \right].
%\nonumber \\
\end{eqnarray}
The electromagnetic form factors $G_0,~G_1$ and $G_2$ above, 
are related by the covariant form factors $F_1,~F_M$ and $F_3$ 
of Eq.~(\ref{current1})~\cite{Gilman:2001yh}: 
\begin{eqnarray}
 G_0 & = & F_1(Q^2) + \frac{2}{3} \eta G_2 (Q^2),\hspace{1ex} ({\rm see\hspace{1ex} below}), \nonumber \\ 
 G_1 & = & F_M(Q^2), \nonumber \\
 G_2 & = & F_1(Q^2) -F_M(Q^2) + (1+ \eta) F_3(Q^2).
\end{eqnarray}

The $\rho$-meson decay constant $f_\rho$ is defined by~\cite{Melo2006,Jaus1991}, 
\begin{equation}
<0|\bar{q}(0)\gamma^\mu q(0)|\phi_\rho(\lambda)>~=~\epsilon^{\mu}_{\lambda} m_\rho f_{\rho}~,
\label{fdecay}
\end{equation}
where $\epsilon^{\mu}_{\lambda}$ is the polarization vector of 
the corresponding $\rho$-meson state $\phi_{\rho}(\lambda)$. 
Note that $f_\rho$ defined above has the mass dimension one, 
the same as the usual definition of the pion decay constant 
(but without a factor $\sqrt{2}$). 
Here, we use the plus-component of the 
electromagnetic current with $\lambda=z$ in the rest frame of the 
$\rho$-meson, and $\epsilon^+_z=1$ with 
$e^\mu_z = (e^+_z,e^-_z,\vec{e}_\perp) = (1,-1,\vec{0})$~\cite{Melo2006}.
The result is independent of the choice of $\lambda$.

We calculate also the $\rho$-meson 
magnetic moment $\mu_\rho$, quadrupole moment $Q_{2 \rho}$ (note the definition below), 
and electromagnetic square charge radius $<r^2_\rho>$.  
They are obtained by the following expressions~\cite{Keister:1993mg}: 
%%%%%%%%%%%%%%%%
\begin{eqnarray}
1 &=& G_0(0), {\rm \hspace{1ex} (charge\hspace{1ex} normailzation)}, 
\label{norm}\\
\mu_\rho &=& G_1(0) = F_M(0), 
\label{murho}\\
Q_{2 \rho} & = & \lim_{Q^2 \to 0} 3 \sqrt{2}\hspace{1ex} \frac{G_2(Q^2)}{Q^2} 
\label{Q2rho}\\
< r^2_\rho >& = & \lim_{Q^2\rightarrow0}\frac{- 6 \left[G_0(Q^2)-1 \right]}{Q^2} 
= - 6 \left. \frac{d G_0(Q^2)}{d Q^2} \right|_{Q^2 = 0}
\label{r2rho}
\end{eqnarray}
%%%%%%%%%%%%%%%

\section{Results}
\label{result}

In the following we present the results for the in-medium $\rho$-meson 
properties calculated in symmetric nuclear matter, namely in-medium electromagnetic 
charge ($G^*_0$), magnetic ($G^*_1$), and quadrupole ($G^*_2$) 
form factors, electromagnetic square charge radius $<r_\rho^{*2}>$, 
and electromagnetic $\rho$-meson decay constant $f^*_\rho$. 
These are calculated by the light-front constituent quark model, 
using the in-medium inputs obtained by the QMC model as already explained.

Before presenting the results, we briefly remind bellow how 
the $\rho$-meson properties are calculated in symmetric nuclear matter.
As explained in section~\ref{QMC}, the light-quark and light-antiquark 
self-energies in symmetric nuclear matter are modified by 
the Lorentz-scalar-isoscalar $\sigma$  and Lorentz-vector-isoscalar $\omega$ mean fields. 
More specifically, in the Hartree approximation, 
the light-quark mass term acquires the attractive Lorentz scalar potential $V^q_\sigma$,  
while the time component of the light-quark (light-antiquark) four-momentum acquires 
the repulsive (attractive) mean field potential $V^q_\omega$. 
Namely, the four-momentum $p^\mu$ of the light-quark (light-antiquark) 
is modified by, 
$p^\mu \to p^{*\,\mu} = p^\mu + V^\mu 
= p^\mu + \, \delta^\mu_0 V^q_{\omega} \hspace{1ex} (p^\mu - \, \delta^\mu_0 V^q_{\omega})$, 
and both the light-quark and light-antiquark masses are modified by  
$m_q \to m_q^* = m_q - V^q_\sigma = m_q - g^q_\sigma \sigma$. 
These mean field potentials are constrained by the nuclear matter 
saturation properties (see Fig.~\ref{fig:qmc} upper panel). 
Then, using the in-medium modified light-quark (light-antiquark) 
properties, as well as the effective $\rho$-meson mass obtained  
in the QMC model (see Fig.~\ref{fig:qmc} lower-right panel), 
we calculate the in-medium $\rho$-meson electromagnetic properties.
Similar approach has already been applied for the studies of pion~\cite{Melo2014}, 
kaon~\cite{Yabusaki:2017sgs} and nucleon~\cite{deAraujo:2017uad} 
properties in symmetric nuclear matter.
In the loop integral appearing in the calculations of electromagnetic form factors 
or the decay constant, we shift the momentum, $k'^\mu=k^\mu + \delta^\mu_0 V^0 \rightarrow k^\mu$, 
and the vector potentials cancel out for the light-quark and light-antiquark systems   
such as pion and $\rho$-meson. 

Furthermore, since the effective $\rho$-meson mass decreases as 
increasing the nuclear matter density in the QMC model (see Fig.~\ref{fig:qmc} right panel), 
the sum of the effective quark masses ($m^*_q + m^*_{\bar{q}}$) 
forming the $\rho$-meson bound state, must be larger than the in-medium 
$\rho$-meson mass ($m^*_\rho$), namely  
the binding energy ($B^*$) to be positive, the same condition as in vacuum, 
to be discussed in detail later. 
We summarize in Table~\ref{tab:meson} some quantities calculated for  
the $\rho$-meson in symmetric nuclear matter.

\begin{table}[htb]
\vspace{3ex}
\begin{center}
\caption{Quantities associated with the in-medium $\rho$-meson properties.  
Effective light-quark mass ($m^*_q$) and in-medium $\rho$-meson mass ($m^*_\rho$) 
are quoted in [GeV], while the $\rho$-meson electromagnetic square charge 
radius $<r_{\rho}^{*2}>$ is in [fm$^2$], 
the electromagnetic decay constant $f^*_\rho$ in [MeV],
the magnetic moment $\mu^*_\rho$ in units of $[e/2m_{\rho}]$, quadrupole moment 
$Q^*_{2 \rho}$ in [fm$^2$], and the momentum $Q^2_{\rm zero}$ 
for $G_0(Q^2_{\rm zero}) = 0$ in [GeV$^2$]. The experimental value  
for $\Gamma_{ee} \equiv\Gamma(\rho^0 \to e^+ e^-) = 7.04 \pm 0.06$ keV in vacuum ($\rho_N = 0$) 
is taken from Ref.~\cite{PDG2014}. 
The numbers given in brackets are the results obtained with the use of the 
density independent regulator mass, $m_R=3.0$~GeV. 
\label{tab:meson}}
%\newline (Obs.:Calculado com 16 pontos de Gauss,cx=0.1~).}
\bigskip
\begin{tabular}{cccccccc}
\hline
\hline
$\rho_N/\rho_0$ &$m^*_q$ &$m^*_{\rho}$ &$<r_{\rho}^{*2}>$ 
&$f^*_\rho $ &$\mu^*_\rho$  &$Q^*_{2 \rho}$ &$Q^2_{\rm zero}$
\\
\hline 
~0    &~0.430    &~0.770  &~0.267     &~153.627    &~2.20  &~-0.0590  &~2.96  \\

~0.01 &~0.427    &~0.767  &~0.255     &~150.451    &~2.20         &~-0.0594      &~2.94   \\
~     &          &        &~(0.270)   &~(153.669)  &~(2.20)       &~(-0.0595)    &~(2.96)  \\
~0.10 &~0.410    &~0.738    &~0.287   &~150.381    &~2.20         &~-0.0636      &~2.74     \\
~     &~          &~         &~(0.296) &~(163.206) &~(2.20)       &~(-0.0639)    &~(2.74)    \\
~0.25 &~0.381    &~0.692    &~0.364   &~121.523    &~2.18         &~-0.0716      &~2.41       \\ 
~     &~         &~         &~(0.352) &~(166.148)  &~(2.19)       &~(-0.0721)    &~(2.43)      \\ 
~0.40 &~0.351    &~0.640    &~0.463   &~121.523    &~2.17         &~-0.08092     &~2.11         \\
~     &~         &~         &~(0.433) &~(174.499)  &~(2.19)       &~(-0.0817)    &~(2.14)        \\
~0.50 &~0.333    &~0.618    &~0.560   &~108.720    &~ 2.18        &~-0.0876      &~1.93           \\
~     &~         &~         &~(0.505) &~(179.325)  &~(2.18)       &~(-0.0884)    &~(1.97)          \\
~0.80 &~0.278    &~0.538    &~1.364   &~78.323     &~2.12         &~-0.1067      &~1.44             \\
~     &~         &~         &~(1.214) &~(178.739)  &~(2.14)       &~(-0.1028)    &~(1.53)            \\
~0.85 &~0.268    &~0.527    &~1.483   &~68.788     &~2.12         &~-0.1128      &~1.43       \\
~     &~         &~         &~(1.353) &~(189.556)  &~(2.12)       &~(-0.1134)    &~(1.48)      \\
~0.90 &~0.260    &~0.514    &~2.013   &~58.696     &~2.10         &~-0.1128      &~1.34         \\
~     &~         &~         &~(1.857) &~(195.042)  &~(2.10)       &~(-0.1152)    &~(1.40)        \\
\hline
Exp.~\cite{PDG2014} &for &$\Gamma_{ee}$ &($\rho_N=0$) &152$\pm 8$     &       &           &        \\
\hline
\hline
\end{tabular}
\end{center}
\vspace{3ex}
\end{table}
%\vspace*{0.5cm}

\begin{figure}[htb]
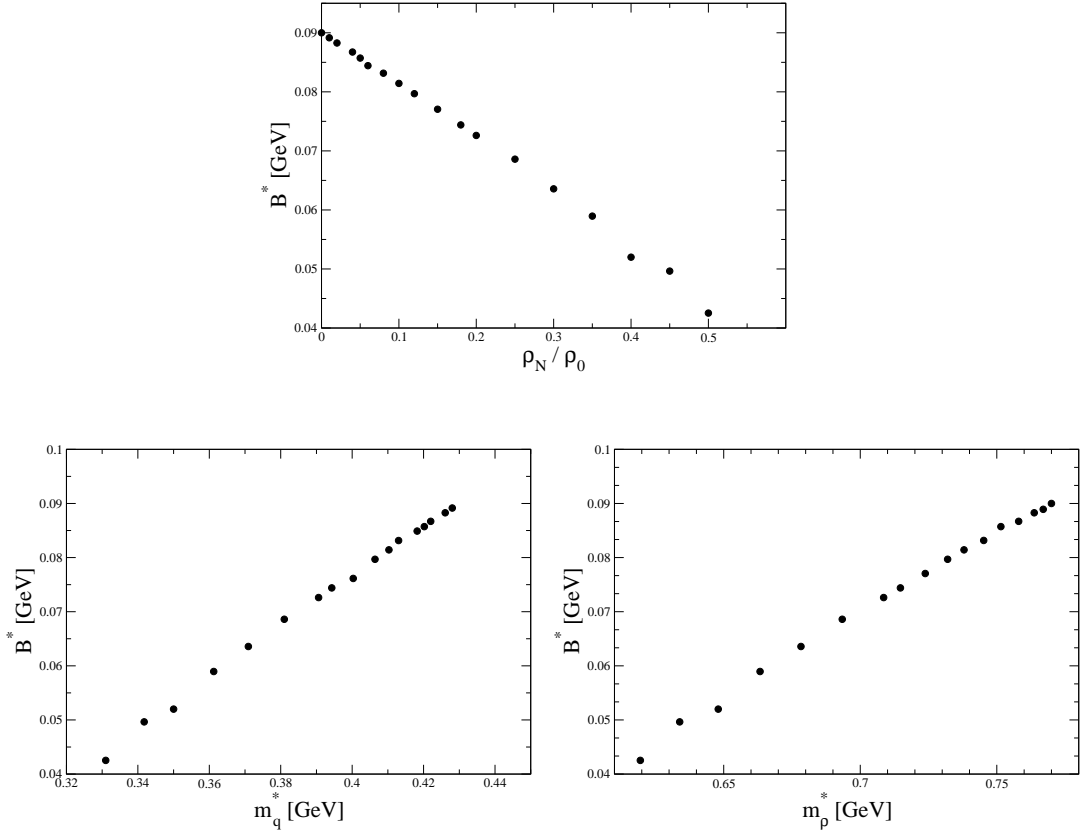

%%\vspace{-2.85cm}
\begin{center}
\epsfig{figure=bdingdensiv1.eps,width=7.0cm}
\vspace{5ex}
\end{center}
\epsfig{figure=bdingmassqv1.eps ,width=7.0cm}  %%%} q2zeromedv1.eps,width=7.5cm} 
\hspace{0.1cm}
\epsfig{figure=bdingrhomassv1.eps,width=7.0cm}
\caption{
\label{bdingdensiv1}
Binding energy $B^*$ [GeV] of the $\rho$-meson versus nuclear matter density 
[$\rho_N/\rho_0$] (upper panel), versus effective light-quark mass $m^*_q$ [GeV] 
(lower-left panel), and versus effective $\rho$-meson mass [GeV] (lower-right panel).
} 
%\vspace{3ex}
\end{figure}

To understand better the bound state nature in the present 
light-front constituent quark model, 
we discuss the binding energy, which should be positive in order to yield  
the bound state for the quark-antiquark composite system. 
The binding energy of the $\rho$-meson in medium $B^*$ is defined by,  
$B^* = m^*_q + m^*_{\bar{q}} - m^*_{\rho}$. 
The binding energy calculated in symmetric nuclear matter is shown 
in Fig.~\ref{bdingdensiv1}, versus the nuclear matter density $\rho_N/\rho_0$ (upper panel),  
versus the effective light-quark mass $m^*_q$ (lower-left panel), and versus 
the $\rho$-meson effective mass $m^*_\rho$ (lower-right panel).
%%%%%%%%%%%%%%%%%%%
%%%%%%%%%%%%
The dependence of the binding energy on the nuclear matter density, effective quark mass 
and effective $\rho$-meson mass, are all nearly linear and smooth.
As the nuclear matter density increases, the binding energy $B^*$  
decreases, and the density beyond about $\rho/\rho_0 = 0.90$  
it becomes negative, and does not yield the $\rho$-meson bound state 
in the present model.
(See also Table~\ref{tab:meson}.)

We comment on the limitation of the light-front constituent quark model 
applied in this study~\cite{Melo1997,Melo2012b,Anace2014}.
Since to yield the bound state in the constituent quark model, 
the binding energy $B$ of the quark-antiquark meson bound state 
must satisfy $B > m_q + m_{\bar{q}}$ in vacuum as well as in medium.
Thus, the constituent quark mass values in vacuum for the 
$\rho$-meson case~\cite{Melo1997,Melo2012b,Anace2014} of $m_q = 430$ MeV and 
pion case of $m_q = 220$ MeV~\cite{Melo2002,Melo2014} are determined  
by the best fit for each case to reproduce the experimental data with 
the regulator mass values $m_R$.
Thus, the present models, although established very well in vacuum, 
cannot study for example $\rho \to \pi \gamma$ transition 
on the same footing even in vacuum.
In this study, we focus on the possible property changes of $\rho$-meson itself in medium, 
and cannot study the reactions involving the mesons with two different constituent 
light quark mass values in the corresponding different mesons consistently, 
as the same reason in vacuum case.

Next, we discuss the $\rho$-meson electromagnetic decay constant in medium, $f^*_\rho$.
The $\rho$-meson decay constant gives direct information on the 
structure of the $\rho$-meson bound state wave function at the origin. 
It is associated with the non-perturbative regime of QCD. 
The $\rho$-meson decay constant is calculated  
via Eq.~(\ref{fdecay}) (see also Ref.~\cite{Melo2006}).
However, the experimental data for $\rho$-meson in vacuum are very scarce~\cite{PDG2014}  
compared with those of the other light mesons such as pion~\cite{Melo2002}. 
The parameters of the model are fitted to the  
empirically extracted $\rho$-meson decay constant from the decay width 
in vacuum~\cite{PDG2014,Clayton2015} (see also Table~\ref{tab:meson}).  
Namely, the light-quark (= light-antiquark) mass of $m_q = 0.430$~GeV, and  
two cases of the regulator masses, density independent case $m_R = 3.0$~GeV, 
and density dependent case, $m_R^* = (m_q^*/m_q) m_R$.
Using these values, the $\rho$-meson decay constant in vacuum obtained 
is $f_\rho = 153.657$~MeV, close to the empirical value of $152 \pm 8$~MeV~\cite{PDG2014}. 
Although we have no clue for the in-medium regulator mass $m_R^*$, or its density 
dependence, as a first trial we assume the density dependence 
$m_R^* = (m_q^*/m_q) m_R$, the same as that of the in-medium light quark 
constituent quark mass, which may be regarded as natural.
We remind that, to calculate the in-medium decay constant $f^*_\rho$, 
we use the in-medium modified polarization vector 
$\epsilon^{* \mu}_{\lambda = z}$ (but $\lambda = z$ is unmodified in medium), 
$\rho$-meson effective mass $m^*_\rho$, and effective quark mass $m^*_q$ in evaluating the  
both sides of Eq.~(\ref{fdecay}).

We show in Fig.~\ref{frho} (left panel) the density dependence of 
the $\rho$-meson decay constant in medium $f^*_\rho$, 
calculated using the density independent regulator mass $m_R = 3.0$ GeV, 
the same as applied in Ref.~\cite{deMelo:2016ynt}. While in the right panel 
we show the result obtained with using the density dependent 
regulator mass, $m_R^* = (m_q^*/m_q) m_R$. 
The solid lines are those interpolated to be able to see easier the density dependence.
The density dependence of $f^*_\rho$ with the density independent $m_R$ is not 
smooth compared to that of the binding energy $B^*$ in Fig.~\ref{bdingdensiv1} (upper panel).
As we will show later, this feature is also noticeable compared with the density 
dependence of the $\rho$-meson electromagnetic form factors and the square charge radius, 
which have smooth density dependence. 
On the other hand, the density dependence of $f^*_\rho$ calculated with 
$m_R^* = (m_q^*/m_q) m_R$ shows smoother density dependence. 
Furthermore, the trends of the density dependence 
of the electromagnetic form factors etc., are similar for the results calculated   
with the two different regulator mass treatments, 
$m_R$ fixed, and $m_R^* = (m_q^*/m_q) m_R$.
In this respect, we may be somehow safe in our predictions for those physical quantities 
to be discussed later. 

Using the two different density dependent $f_\rho^*$, 
we calculate the $\rho^0$-meson decay width to $e^+ e^-$, 
with the formula, 
%%%%%%%%%%%%%%%%
\begin{equation}
\Gamma^*(\rho^0 \to e^+ e^-) = \frac{4 \pi}{3} \frac{\alpha_e^2}{m^*_\rho} {f^{*\,2}_\rho}, 
\label{Gamma}
\end{equation}
%%%%%%%%%%%%%%
where $\alpha_e$ is the electromagnetic fine structure constant.
Note that, the decrease (increase) of $f^*_\rho$ as increasing the nuclear matter density  
is the similar (opposite) behavior to that of the pion decay constant 
$f^*_\pi$~\cite{Melo2014}, which decreases in both the space component 
and the time component~\cite{Kirchbach:1997rk} as increasing 
the nuclear matter density.

\begin{figure}[tb]
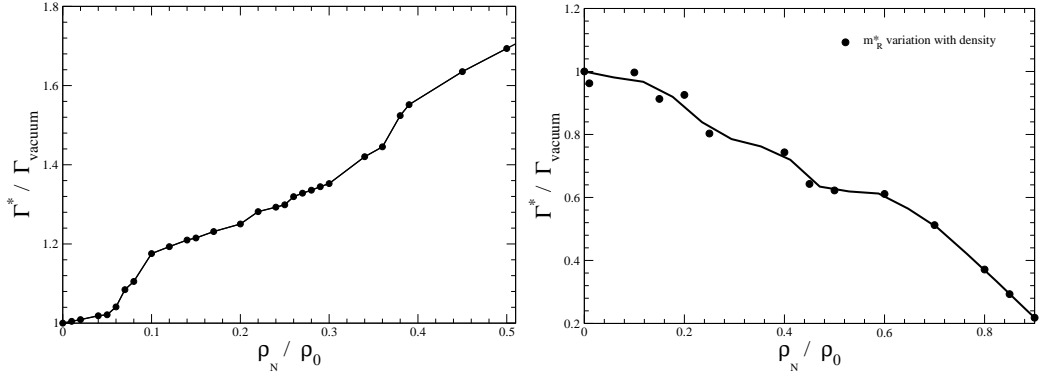

%\vspace{3ex}
%%\vspace{-2.85cm}
\epsfig{figure=ratioswidthv1.eps,width=6.8cm}
\epsfig{figure=ratioswidthv3.eps,width=6.8cm}    
\caption
{Density dependence of the decay width ratios 
$\Gamma^*(\rho^0 \to e^+ e^-)/\Gamma(\rho^0 \to e^+ e^-)$, 
calculated with the fixed $m_R$ (left panel), 
and with $m_R^* = (m_q^*/m_q) m_R$ (right panel), 
where in vacuum $\Gamma(\rho^0 \to e^+ e^-) = 7.04 \pm 0.06$ keV~\cite{PDG2014}.
%\label{frho}
}
\label{width}
\end{figure}

In Fig.~\ref{width} we show the density dependence of the decay width ratio 
to the vacuum, $\Gamma^*(\rho^0 \to e^+ e^-)/\Gamma(\rho^0 \to e^+ e^-)$, 
with the density independent (left panel) and density dependent (right panel) regulator mass, 
where $\Gamma(\rho^0 \to e^+ e^-) = 7.04 \pm 0.06$ keV~\cite{PDG2014} in vacuum.
%\vspace{0.50cm}
%%%%%%%%%%%%%%%%%%%
\begin{figure}[tb]
%\vspace{3ex}
%%\vspace{-2.85cm}
\epsfig{figure=fdecayv2.eps,width=7.0cm}
\epsfig{figure=fdecaymrvarv3.eps,width=7.0cm}
\caption
{Density dependence of the $\rho$-meson decay constant $f^*_\rho$ calculated with 
the fixed $m_R$ (left panel), 
and that calculated with $m_R^* = (m_q^*/m_q) m_R$~(right panel). 
%% $of the decay width ratio, $\Gamma^*(\rho^0 \to e^+ e^-)/\Gamma(\rho^0 \to e^+ e^-)$ (right panel), 
%% where $\Gamma(\rho^0 \to e^+ e^-) = 7.04 \pm 0.06$ keV~\cite{PDG2014} in vacuum.
\label{frho}
}
%\label{fidecay2}
\end{figure}

The results for the in-medium decay width, $\rho^0 \to e^+ e^-$,  
show opposite density dependence, increase (left panel) and 
decrease (right panel) as increasing the nuclear matter density, 
reflecting respectively the density dependence of $f_\rho^*$ 
(left panel) and (right panel) directly shown in Fig.~\ref{frho}.
This density dependence of the $\rho^0 \to e^+ e^-$ in nuclear 
medium (in nuclei) may be indirectly observed in experiment, 
and can give important information to draw a more solid conclusion on 
the density dependence of $f_\rho^*$.  
 
\begin{figure}[tb]
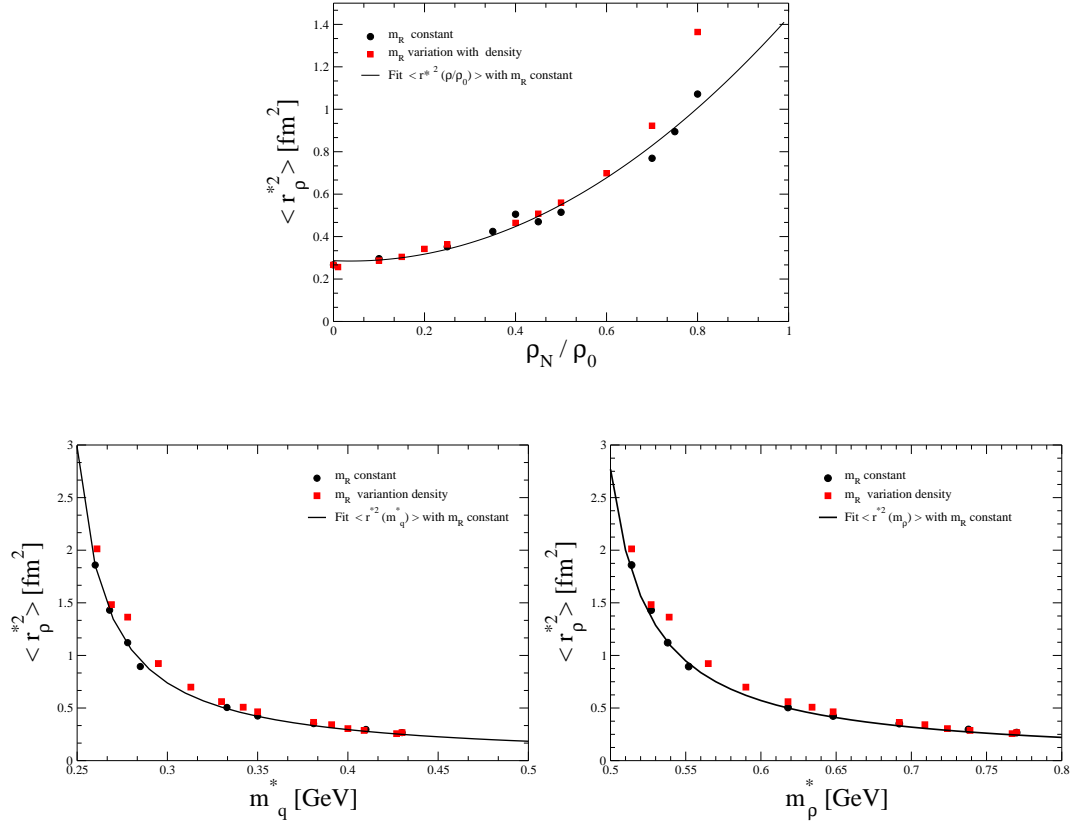

\begin{center}
\epsfig{figure=radiusdensiv3.eps,width=7.0cm}  %% radiusdensiv2.eps,width=7.0cm}
\vspace{5ex}
\end{center}
\epsfig{figure=radiusmqmrvarv3.eps,width=7.0cm}      %% radiusmqv1.eps ,width=5.0cm}
\epsfig{figure=radiusmrhomrvarv3.eps,width=7.0cm}    %% radiusv1,width=7.0cm}
\caption{\label{radius} $\rho$-meson electromagnetic square charge radius  
as well as the fitted curves, 
versus nuclear matter density [$\rho_N/\rho_0$] (upper panel), 
versus effective quark mass $m^*_q$ (lower-left panel), 
and versus effective $\rho$-meson mass $m^*_\rho$.
The filled circles and filled squares are respectively the results obtained with 
the density independent and density dependent regulator mass.
\vspace{3ex}
}
\label{fitradius}
\end{figure}

We note that, the behavior of increasing $f^*_\rho$ with the increase of 
the square charge radius $<r^{*\,2}_\rho>$   
which will be shown later in Fig.~\ref{radius}, may be consistent 
with the correlation observed in vacuum in Ref.~\cite{Krutov2016}.
Thus, we are in a difficult position to draw any 
conclusions on the density dependence of $f_\rho^*$ at this moment.
We need to wait some useful information from experiment.  

Now we discuss the main results of this article, $\rho$-meson electromagnetic 
form factors in symmetric nuclear matter.
As we mentioned already the results given below are calculated with the 
density dependent regulator mass, $m_R^* = (m_q^*/m_q) m_R$.

\begin{figure}[t]
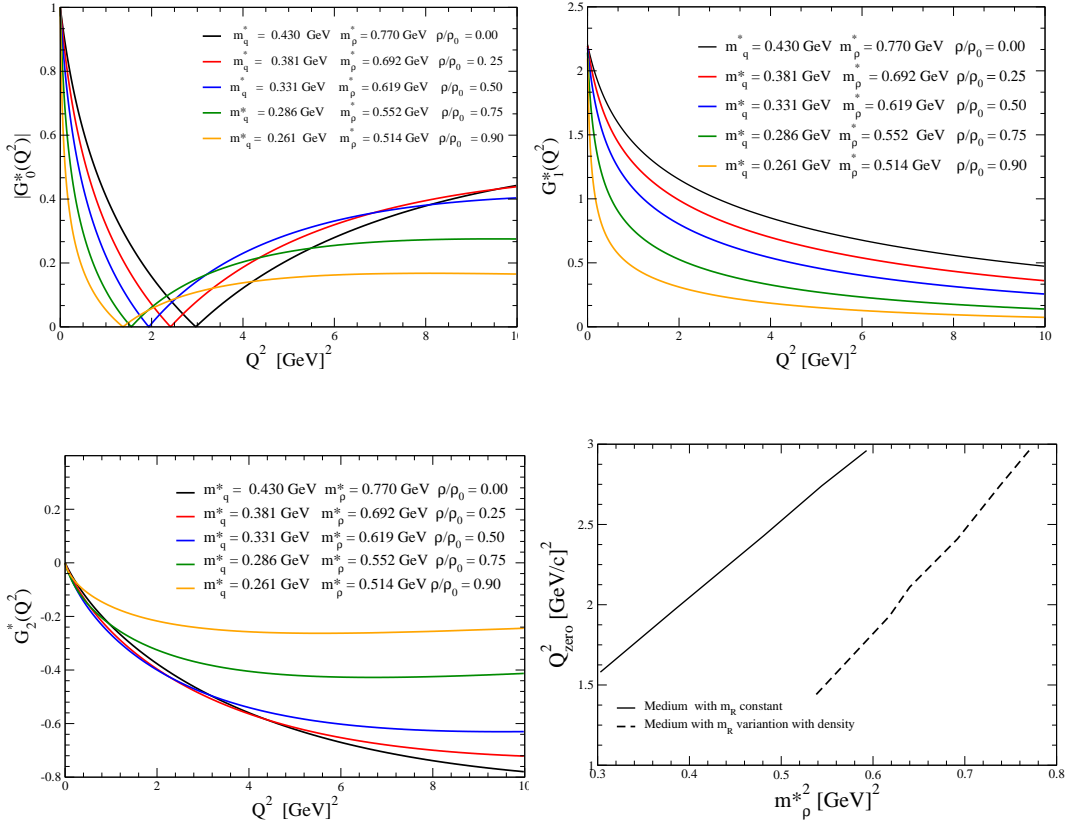

%%  \hspace{-0.85cm}
\epsfig{figure=g0mediumv3.eps,width=6.9cm}
\epsfig{figure=g1mediumv3.eps,width=6.9
cm}
\vspace{2ex}
\\
\epsfig{figure=g2mediumv3.eps,width=7.0cm}
\epsfig{figure=q2zeromediunv3.eps,width=7.0cm}
\caption{
\label{figg0}
$\rho$-meson electromagnetic charge $|G_0|$ (upper left panel),
magnetic $G_1$ (upper-right panel), and  
quadrupole $G_2$ (lower-left panel) form factors for several densities, 
and the $Q^2_{\rm zero}$ for $G_0(Q^2_{\rm zero}) = 0$ versus 
$m^{* 2}_\rho$ (lower-right panel) in symmetric matter.
}
\end{figure}

In Fig.~\ref{figg0} we show the $\rho$-meson 
charge $|G_0(Q^2)|$ (upper-left panel), 
magnetic $G_1(Q^2)$ (upper-right panel),
quadrupole $G_2(Q^2)$ (lower-left panel) form factors for several 
nuclear matter densities, and the zero, $Q_{\rm zero}^2 = - q_{\rm zero}^2$,  
to give $G_0(Q_{\rm zero}^2) = 0$ for the charge form factor, 
versus the effective $\rho$-meson mass $m^{*\,2}_\rho$ (lower-right panel).
For the zero, $Q_{\rm zero}^2$, we also show that the result obtained with 
the density independent $m_R$.
%%%%%%%%%%%%%%%%%% 
%%%%%%%%%%%%

The $\rho$-meson electromagnetic form factors in symmetric nuclear matter 
~$G^*_0,~G^*_1$ and $G^*_2$ are strongly modified as increasing the nuclear matter density. 
The modification of $|G^*_0|$ shows two distinct features: (i) faster decrease near $Q^2 = 0$ 
relative to that in vacuum, which implies the increase of the charge radius in symmetric nuclear matter, 
and (ii) the position of $Q^2_{\rm zero}$ decreases as increasing the nuclear matter density 
almost linearly for the two cases of the regulator mass, the density dependent and independent. 
The present model has $Q^2_{\rm zero} \simeq 3$ GeV$^2$ in vacuum~\cite{Melo1997}, where 
in the literature the values in vacuum are spread in the region, 
$3\, {\rm GeV^2} < Q^2_{\rm zero} < 5\, {\rm GeV^2}$
~\cite{Cardarelli1995,Choi2004,Bhagwat2008,Roberts2011,deMelo2018b2,Adamuscin2007,Bakker2002}. %% ,Melo2017v1}.

The fact that $G_0$ has the zero, is similar to the spin-one deuteron case, 
a composite system of spin-1/2 particles of  
proton and neutron~\cite{Gilman:2001yh,Buchmann1989,Carlson1989}.
%%The $Q^2_{\rm zero}$ values versus $m_\rho^{*\,2}$ show nearly the 
%%linear dependence for both the density independent $m_R$ and 
%%density dependent $m_R^* = (m_q^*/m_q) m_R$.
%Other form factors $G^*_1$ and $G^*_2$ also decrease faster than those 
%%corresponding form factors in vacuum as increasing the nuclear matter density.

The $Q^2_{\rm zero}$ versus $m^{* 2}_\rho$ shown in the lower-right panel 
in Fig.~\ref{figg0} is very interesting and may be noticeable.
The use of the two different regulator mass $m_R$ and $m_R^* = (m_q^*/m_q) m_R$, 
both show the nearly linear dependence.
%%The position of $Q^2_{\rm zero}$ in terms of $m^{* 2}_\rho$ can be expressed by, 
%%\begin{equation}
%%Q^2_{\rm zero} \simeq 5.0 \times m^{* 2}_\rho + c,  
%%\end{equation}
%%with the constant $c \simeq 0$.  

For the $\rho$-meson magnetic moment in medium $\mu^*_\rho = G^*_1(0)$, 
the medium modification is the very small quenching 
(see Table~\ref{tab:meson}). In vacuum, $\mu_\rho = 2.20$ in the present 
model~\cite{Melo1997,Clayton2015} (and $\mu_\rho = 2.16$ in Ref.~\cite{Krutov:2018mbu}), 
while $\mu^*_\rho = 2.10$ at $\rho_N/\rho_0 = 0.90$. 
This is in contrast with the nucleon case, which has been demonstrated to 
be enhanced as increasing the nuclear matter density~\cite{deAraujo:2017uad,Lu:1998tn}.
This is probably due to the difference in the Lorentz structure between the spin-1  
and spin-1/2 particles (see Eqs.~(\ref{current1}) and~(\ref{murho})). 

The $\rho$-meson quadrupole moment $Q_{2 \rho}$ calculated via Eq.~(\ref{Q2rho}), 
is also sensitive to the effects of the nuclear medium (see Table~\ref{tab:meson}).
The quadrupole form factor $G^*_2$ shown in Fig.~\ref{figg0} (lower-left panel), 
changes behavior from quenching to enhancement for the densities larger 
than $\rho/\rho_0=0.75$.  
The enhancement of the quadrupole moment $Q^*_{2 \rho}$  
as increasing the nuclear matter density (see Table~\ref{tab:meson}), 
means that the $\rho$-meson electromagnetic charge distribution 
deviates more from spherical symmetry in symmetric nuclear matter.

Next, we discuss the $\rho$-meson electromagnetic square charge radius  
in symmetric nuclear matter $< r^{* \,2}_\rho>$ calculated with Eq.~(\ref{r2rho}). 
The results are shown in Fig.~\ref{fitradius}, versus nuclear matter density (upper panel), 
versus the effective quark mass $m^*_q$ (lower-left panel), 
and versus the effective $\rho$-meson mass $m^*_\rho$ (lower-right panel).
These are shown for the two cases, with the density independent (filled circles) 
and density dependent (filled squares) regulator mass. 
The solid line in each panel is a line obtained by fitting to the points calculated, 
for helping to see easier.
%%%%%%%%%%%%%%%%%%%
%\vspace{0.5cm}
%%%%%%%%%%%%%

%As for the upper panel in Fig.~\ref{fitradius} 
%the solid line is obtained by the following expression  
%for the range $0 \leq y \equiv \rho_N/\rho_0 \leq 0.9$, 
%%%%%%%%%%%%%%%%Gilman:2001yh
%\begin{equation}
%<~r_\rho^{*2}(y)~>~=~ C_0 y^2 + C_1 y + C_2  = 1.24173 y^2 -0.0943938 y + 0.286568,  
%\label{fitden}
%\end{equation}
%%%%%%%%%%%%%%%
%where $C_{0,1,2}$ above are all in [fm$^2$]. 
%$C_0=1.24173$ [fm$^2$], $C_1=-0.0943938$ [fm$^2$], and
%$C_2=0.286568$ [fm$^2$]. 

%For the lower-left panel in Fig.~\ref{fitradius}, the fit function   
%is obtained for the region  
%$0.2\, {\rm GeV} \leq m^*_q \leq 0.5\, {\rm GeV}$ as,  
%%%%%%%%%%%%%%%%
%\begin{equation}
%<~r_\rho^{*2}(m^*_q)~>~=
%~\frac{a_0}{m^*_q-a_1}~=~
%~\frac{0.049064}{m^*_q-0.233531},
%\label{fitmq}
%\end{equation}
%%%%%%%%%%%%%%
%with~$a_0$ in [GeV\,fm$^2$] and $a_1$ in [GeV].

%For the lower-right panel in Fig.~\ref{fitradius}, 
%the fit function for the region 
%$0.50\, {\rm GeV} \leq m^*_\rho \leq 0.80\, {\rm GeV}$ obtained is,  
%\begin{equation}
%% F_1(m^*_\rho ) = 
%<~r_\rho^{*2}(m^*_{\rho})~>~=~\frac{b_0}{m^*_{\rho}-b_1}~
%~=~\frac{0.0720}{m^*_{\rho}-0.474},
%\label{fitmrho}
%\end{equation}
%with $b_0$ in [GeV\,fm$^2$] and $b_1$ in [GeV]. 

Three features shown in Fig.~\ref{fitradius} for $< r^{*\,2}_\rho >$ can be understood as follows.
As the nuclear matter density increases, the charge form factor $G^*_0$ decreases faster 
near around $Q^2 = 0$ as shown in Fig.~\ref{figg0} upper-left panel, 
and the derivative with respective to $Q^2$ becomes negatively larger at $Q^2 = 0$ 
(see Eq.~(\ref{r2rho})) to yield larger $< r^{*\,2}_\rho >$.
As increasing the nuclear matter density, both $m^*_q$ and $m^*_\rho$ decrease 
as shown in Fig.~\ref{fig:qmc} lower-left and lower-right panels, respectively. 
Furthermore, as the nuclear matter density increases, $m^*_q$ and  
$m^*_\rho$ as well as the binding energy $B^*$ decrease as 
shown in Fig.~\ref{bdingdensiv1}.
The decrease in the binding energy yields a looser bound state, 
thus resulting in the increase of $< r^{*\,2}_\rho >$.

\section{Summary and conclusion}
\label{conclusion}

We have studied the $\rho$-meson electromagnetic properties in symmetric nuclear matter 
with a light-front constituent quark model using the in-medium inputs 
calculated by the quark-meson coupling model.
Similar approach was already applied in the studies 
of the pion, kaon, and nucleon properties in symmetric nuclear matter.

In this study we have applied a density dependent regulator mass, 
while in our previous study we used the density independent 
regulator mass. The results obtained in this extended study 
predict the similar density dependence of the $\rho$-meson 
electromagnetic properties, except for the density dependence 
of the $\rho$-meson electromagnetic decay constant. 
The present result shows the decrease of the $\rho$-meson 
electromagnetic decay constant, opposite behavior to the initial 
study obtained with the density independent regulator mass.
Thus we need to wait relevant experimental data from which we 
can draw a more definite conclusion on the density dependence of 
the $\rho$-meson decay constant.

Except for the $\rho$-meson decay constant, we first predict 
the $\rho$-meson electric, magnetic, and quadrupole form factors to 
vary faster in symmetric nuclear matter as increasing the nuclear matter density 
than those in vacuum, versus the (negative of) four-momentum transfer squared.

Second, we predict that, as increasing the nuclear matter density, 
the $\rho$-meson charge radius and modulus 
of the quadrupole moment increase (enhanced), 
while the magnetic moment is slightly quenched.
The quenching of the magnetic moment shows the opposite behavior compared with 
that of the spin-1/2 Dirac particle (known to be enhanced), 
probably due to the difference in the Lorentz structure.
The enhancement of the quadrupole moment in symmetric nuclear matter 
means that the $\rho$-meson charge distribution is more deviate 
from a spherical symmetric distribution in symmetric nuclear matter.
Furthermore, we predict the value of ``zero'', 
the value of the (negative of) four-momentum transfer squared
to cross zero of the $\rho$-meson charge form factor (positive value), 
becomes smaller as increasing the nuclear matter density. 
This feature is also the same as that obtained with the use 
of the density independent regulator mass.

Although the present situation does not allow us to have many experimental data 
in vacuum as well as in nuclear medium, we hope further advances in experiments 
will provide us with more relevant data on the $\rho$-meson properties in vacuum 
and in a nuclear medium (a nucleus).
In particular information on the density dependence of the 
$\rho$-meson electromagnetic decay constant, such as the corresponding 
decay width in-medium even though indirect manner, may be very useful.

For a future prospect, we need to study different scheme of the 
regularization, and/or the density dependence of the regulator mass,  
and also plan to extend the similar approach to 
study the in-medium properties of $K$-, $D$-, $K^*$- 
$D^*$- and $B$-mesons.  
\\ 

{\bf Acknowledgements}
\\

This work was partially supported by the Funda\c c\~ao de Amparo \`a Pesquisa do Estado de
S\~ao Paulo (FAPESP),~Brazil, No.~2015/16295-5 (JPBCM), and No.~2015/17234-0 (KT), 
~and Conselho Nacional de Desenvolvimento 
Cient\'ifico e Tecnol\'ogico~(CNPq), Brazil, No. 401322/2014-9 (JPBCM), 
No. 400826/2014-3 (KT), No. 308025/2015-6 (JPBCM), and No. 308088/2015-8 (KT).
This work was part of the projects, 
Instituto Nacional de Ci\^{e}ncia e Tecnologia - Nuclear Physics and Applications 
(INCT-FNA), Brazil, No. 464898/2014-5, and FAPESP Tem\'{a}tico, No. 2017/05660-0.

\end{document}